\documentclass[12pt]{elsarticle}
\usepackage{graphicx}
\usepackage{bm}
\usepackage{amsmath,amssymb,amsthm,dsfont,bm,soul}
\usepackage{color}
\usepackage{wasysym}
\usepackage{ulem}
\usepackage{appendix}
\newcommand{\AL}[1]{\textcolor{black}{#1}}
\newcommand{\LA}[1]{\textcolor{black}{#1}}
\usepackage[dvipsnames]{xcolor}
\usepackage[english]{babel} 
\usepackage[utf8]{inputenc} 
\usepackage{hyperref}
\usepackage{cancel}

\begin{document}
\begin{frontmatter}
\title{Inequalities for complementarity in observed statistics}
\author{Elisa Masa}
\author{Laura Ares}
\ead{laurares@ucm.es}
\author{Alfredo Luis}
\ead{alluis@ucm.es}
\affiliation{Organization={Departamento de \'{O}ptica, Facultad de Ciencias F\'{i}sicas, Universidad Complutense},
            addressline={Plaza de ciencias, 1}, 
            city={Madrid},
            postcode={28040}, 
            country={Spain}}

\begin{abstract}
We provide an analysis  of  complementarity via a suitably designed  classical model that leads to a set of inequalities that can be tested by means of unsharp measurements. We show that, if the measured statistics does not fulfill the inequalities it is equivalent to the lack of a joint distribution for the incompatible observables. This is illustrated by path-interference duality in a Young interferometer.
\end{abstract}

\end{frontmatter}

\section{Introduction}
The concept of complementarity has been deeply debated from its very formalization \cite{NB28, WMM02,LB90} which is still aim of discussion from a qualitative but also quantitative point of view \cite{JB10,XQ20,YC21}. In addition, it is one of the basic ingredients of Bell analysis, along with entanglement\cite{JB64,CH74,NV62,MA84,LA87,MC88,AK00,HP04,AM08,TN11,AK14,AK19,AF82,RS08,AR15,BKO16,MAL20} and it plays an essential role in new quantum technologies \cite{MK09,AK21}.

\bigskip

In this work we look for inequalities involving observable mean values and correlations that may disclose complementarity as a nonclassical effect incompatible with classical statistical models. To this end, we propose a practical feasible scheme based on the joint measurement of two compatible observables that can be regarded as fuzzy or noisy counterparts of two system incompatible observables. The key point is to design the joint measurement such that its statistics provide complete information to retrieve the exact statistics of the two system observables after a suitable data inversion \cite{MAL20,MM89,WMM02,WMM14,PB87, AL16a,AL16b,LM17,GBAL18}. 

\bigskip

We derive inequalities for the observed statistics that must be fulfilled if the system and detection process satisfy some natural requirements of the classical theory, so their violations becomes a clear quantum signature. Furthermore we show that this violation is equivalent to a lack of a noise-free joint distribution for these incompatible observables \cite{AF82,RS08,AR15,BKO16,MAL20}. All this is then illustrated by path-interference duality in a Young interferometer.

\bigskip

\section{Methods}
Let us present the details of the observation scheme and data inversion that will provide the key ingredients for the ensuing analyses. In Sec. IV we provide a suitable physical realization of the whole scheme. 

\subsection{System and joint measurement } 
Let us consider a two-dimensional system in some arbitrary state $\rho$ 
\begin{equation}
\label{rho}
\rho= \frac{1}{2} \left  ( \sigma_0 +  \bm{s} \cdot  \bm{\sigma}  \right )   , 
\end{equation}
being $\bm{\sigma}$ the Pauli matrices, $\sigma_0$ is the identity, and $\bm{s}$ a  real vectors with $|\bm{s}| \leq 1$. The two incompatible observables to be considered will be represented by the Pauli matrices $\sigma_X$ and $\sigma_Z$. 

\bigskip

The noisy simultaneous measurement  will proceed following standard procedures. The observed system is coupled to auxiliary degrees of freedom after which two compatible observables are measured, say $X$ and $Z$, in the combined system-auxiliary space. For the sake of simplicity we will assume that the results of such pair of measurements are dichotomic and labelled by the variables $x,z= \pm 1$, in accordance with the spectra of $\sigma_X$ and $\sigma_Z$. Let us call the observed statistics $\tilde{p} (x,z)$. As it can be easily seen by a Taylor series in powers of $x$ and $z$, the most general form for $\tilde{p} (x,z)$ is 
\begin{equation}
    \tilde{p} (x,z) = \frac{1}{4} \left ( 1+ x \overline{x} + z \overline{z} + xz  \overline{xz} \right ) ,
\end{equation}
where naturally $\overline{x}$, $\overline{z}$, and $\overline{xz}$ are the corresponding mean values and correlations of $X$ and $Z$, 
\begin{equation}
    \overline{x} = \sum_{x,z = \pm 1} x \tilde{p} (x,z) , \quad \overline{z} = \sum_{x,z = \pm 1} z \tilde{p} (x,z) ,
\end{equation}
and 
\begin{equation}
    \overline{xz} = \sum_{x,z = \pm 1} x z \tilde{p} (x,z) ,
\end{equation}
and they are so that $ \tilde{p} (x,z) \geq 0$ for all $x,z$.

\bigskip

In order to link $\overline{x}$, $\overline{z}$, and $\overline{xz}$ with system observables we further assume that 
\begin{equation}
\label{mxmz}
    \overline{x} = \gamma_X \mathrm{tr} \left ( \rho  \sigma_X \right ) , \qquad \overline{z} = \gamma_Z \mathrm{tr} \left ( \rho  \sigma_Z \right ) ,
\end{equation}
where $\gamma_X$ and $\gamma_Z$ are two real factors that express the noisy character of the joint observation. Note that for dichotomic variables the variance is determined by the mean value $\Delta^2 x = 1 - \overline{x}^2$, so $\Delta^2 x$  is always larger than $\Delta^2 \sigma_X$ , the lesser $\gamma_X^2$ the larger the difference. For definiteness and without loss of generality we consider $ 1 \geq \gamma_X,\gamma_Z > 0$. 

\bigskip

Likewise, invoking quantum linearity we further assume that  
\begin{equation}
\label{mxz}
    \overline{xz} = \gamma_{XZ} \mathrm{tr} \left ( \rho  \sigma_n \right ) ,
\end{equation}
where $\sigma_n = \bm{\sigma} \cdot \bm{n}$ being $\bm{n}$  a real unit vector, and again we assume $ 1 \geq \gamma_{XZ} \geq 0$ without loss of generality, since the sign of $\overline{xz}$ can be fully absorbed in $\bm{n}$. Notice that for the observed variables there is no ambiguity regarding the meaning of $\overline{xz}$ which is just the mean value of the commuting product $XZ$.

The $\gamma$ parameters are not independent, since the joint distribution $\tilde{p} (x,z)$ must be nonnegative in all cases. The particular form of the constraint depends on $\sigma_n$. For example, if $\sigma_n = \sigma_y$ then $\gamma_X^2+\gamma_Z^2+\gamma_{XZ}^2 \leq 1$, which is actually the case of the Young example to be examined in detail below, while if $\sigma_n = \sigma_x$ we would have $( \gamma_X + \gamma_{XZ})^2+\gamma_Z^2 \leq 1$.

\bigskip

It is worth noting that these $\gamma$ relations bear some resemblance with well-known relations connecting path, interference, and entanglement, as presented for example in Ref. \cite{JB10}. In this regard, the observable $X$ may represent interference, \AL{while $Z$ may represent} path. Moreover, the correlation $XZ$ is actually at the heart of a nonclassical behavior as far as its quantum version $\sigma_X \sigma_Z$ is rather undefined or ambiguous. Nevertheless, that analogy is not complete in the sense that our model involves a single system so there is no entanglement at work. Moreover, the factors $\gamma$ are not system observables but parameters characterizing the measurement process granting that $X$ and $Z$ are indeed compatible observables as assumed.

\bigskip

\section{Classical variable model}



Complementarity has to do with the statistics of two observables, in our case $\sigma_X$ and $\sigma_Z$. The main result of classical physics that the quantum theory does not admit is the existence of a joint probability distribution for the values $x^\prime$ and $z^\prime$ that $\sigma_X$ and $\sigma_Z$ can take. According to our objectives, this is the main assumption of our classical-like model: the existence of such distribution, $p_\Lambda(\lambda)$ for  $\lambda =(x^\prime,z^\prime)$. Moreover we assume that the values $(x^\prime,z^\prime)$ that $\sigma_X$ and $\sigma_Z$ can take are the results we \AL{would} obtain when measuring them alone, this is $x^\prime,z^\prime =\pm 1$. This is to say that all elements of our model are observable.

\bigskip

\subsection{Measurement independence}
As a further assumption derived form a classical-theory framework we consider measurement independence or separability, understood as the idea that the \AL{observed} statistics of $X$ does not depend on the \AL{joint observation} $Z$, \AL{and vice versa}. Say, referring to a more familiar situation, there is no necessary influence of the measurement of momentum on the measurement of position. So we say that $\tilde{p} (x, z)$ is separable provided that 
\begin{equation}
\label{Bt}
    \tilde{p} (x,z) = \int d\lambda p_X(x|\lambda) p_Z(z|\lambda) p_\Lambda(\lambda) ,
\end{equation}
where \AL{$p_X(x|\lambda)$ and $p_Z(z|\lambda)$} are conditional probabilities and naturally $p_X (x|\lambda)$, $p_Z(z|\lambda)$, and $p_\Lambda (\lambda)$, are assumed to be real, nonnegative, and normalized in $x$, $z$ and $\lambda$, respectively.

\bigskip

\subsection{Statistical independence}
In this spirit we may also naturally assume that the conditional probabilities take the form $p_X (x|\lambda)= p_X (x|x^\prime)$  and $p_Z(z|\lambda)=p_Z(z|z^\prime)$, so that the statistics of $X$ is derived exclusively in terms of the statistics of $\sigma_X$, and equivalently, the statistics of $Z$ is derived exclusively in terms of the statistics of $\sigma_Z$. This is specially clear if we consider the individual measurements with $X= \sigma_X$ for example, this is $\gamma_X=1$ and $\gamma_Z = \gamma_{XZ}=0$. To provide suitable conditional probabilities, $p_X (x|x^\prime)$ and $p_Z(z|z^\prime)$ we focus on the observed marginal statistics for $X$ and $Z$ 
\begin{equation}
\label{marg}
    \tilde{p}_X (x) = \sum_{z = \pm 1} \tilde{p} (x,z) =  \frac{1}{2} \left [ 1+ \gamma_X x \mathrm{tr} \left ( \rho  \sigma_X \right ) \right ] ,
\end{equation}
and similarly for $Z$, to propose the following conditional probabilities 
\begin{equation}
\label{fac1}
   p_X (x|x^\prime )= \frac{1}{2} \left ( 1+ \gamma_X x x^\prime \right ), 
\end{equation}
and likewise
\begin{equation}
\label{fac2}
    p_Z (z|z^\prime) =\frac{1}{2} \left ( 1+ \gamma_Z z z^\prime \right ).
\end{equation}

\subsection{State representation}

After all these assumptions, we will consider a state probability distribution $p_\Lambda (x^\prime, z^\prime)$  so that
\begin{equation}
\label{plambda}
    \tilde{p} (x,z) =\sum_{x^\prime, z^\prime = \pm 1}  p_X (x|x^\prime) p_Z (z|z^\prime) p_\Lambda(x^\prime,z^\prime) ,
\end{equation}
along with Eqs. (\ref{fac1}) and (\ref{fac2}). 
The simplicity of this model allows us to find a unique solution for $p_\Lambda (x^\prime, z^\prime)$, resorting again to a Taylor series on $x^\prime,z^\prime$, which is 
\begin{equation}
\label{plp}
p_\Lambda (x^\prime, z^\prime) = \frac{1}{4} \left ( 1+ x^\prime \frac{\overline{x} }{\gamma_X}+ z^\prime \frac{\overline{z}}{\gamma_Z} + x^\prime z^\prime \frac{\overline{xz}}{\gamma_X \gamma_Z} \right ).
\end{equation}
This is the most general distribution compatible with the natural physical assumptions within a classical theory and the experimental data. The analysis is then reduced to the question of when $p_\Lambda(x^\prime,z^\prime)$ is a probability distribution or not.  

\bigskip

\subsection{Complementarity inequalities}

The distribution $p_\Lambda (x^\prime, z^\prime)$ is a {\it bona fide} probability distribution provided that $p_\Lambda (x^\prime, z^\prime) \geq 0$ for all $x^\prime, z^\prime$ leading to the following family of four inequalities: 
\begin{eqnarray}
\label{Bi}
     & 1+ \frac{\overline{x} }{\gamma_X}+ \frac{\overline{z}}{\gamma_Z} + \frac{\overline{xz}}{\gamma_X \gamma_Z} \geq 0 , \quad  1 - \frac{\overline{x} }{\gamma_X} -
\frac{\overline{z}}{\gamma_Z} + \frac{\overline{xz}}{\gamma_X \gamma_Z} \geq 0 , &  \nonumber \\ & & \\
     &  1 - \frac{\overline{x} }{\gamma_X}+ \frac{\overline{z}}{\gamma_Z} - \frac{\overline{xz}}{\gamma_X \gamma_Z} \geq 0 , \quad  1+ \frac{\overline{x} }{\gamma_X}- \frac{\overline{z}}{\gamma_Z} - \frac{\overline{xz}}{\gamma_X \gamma_Z} \geq 0 
      .&  \nonumber 
\end{eqnarray}
These can be regarded as conditions on the observed mean values $\overline{x}$, $\overline{z}$ and correlations $\overline{xz}$ to be consistent with the classical model. 

\bigskip

Naturally, these four relations can be summarized in a single one, that is the most strict among them. Using that the minimum of two real number $a$,$b$ is 
\begin{equation}
\label{min}
    \min (a,b) = \frac{a+b-|a-b|}{2} ,
\end{equation}
we have that he minimum of the first and second rows in Eq. (\ref{Bi}) are, respectively,
\begin{equation}
    1+ \frac{\overline{xz}}{\gamma_X \gamma_Z} - \left | \frac{\overline{x} }{\gamma_X}+ \frac{\overline{z}}{\gamma_Z} \right | \geq 0 , \quad
      1 - \frac{\overline{xz}}{\gamma_X \gamma_Z} - \left | \frac{\overline{x} }{\gamma_X}-  \frac{\overline{z}}{\gamma_Z} \right | \geq 0 ,
\end{equation}
so we get the final form for our inequalities as
\begin{equation}
\label{tBi}
    1 - \left | \frac{\overline{x} }{\gamma_X} - \frac{\overline{z}}{\gamma_Z} \right | \geq \frac{\overline{xz}}{\gamma_X \gamma_Z} 
    \geq  \left | \frac{\overline{x} }{\gamma_X} +  \frac{\overline{z}}{\gamma_Z} \right |  - 1.
\end{equation}
Finally, they can be expressed in terms of the parameters of the density matrix $\rho$ in Eq. (\ref{rho})
\begin{equation}
\  1 - \left | s_X - s_Z \right | \geq \frac{\gamma_{XZ}}{\gamma_X \gamma_Z} s_n
    \geq  \left | s_X + s_Z \right |  - 1.
\end{equation}
being $s_j =  \mathrm{tr} \left ( \rho  \sigma_j \right )$, $j=x,z,n$.

\bigskip

Thanks to this formulation of observed complementarity we can rigorously demonstrate a result already anticipated in Ref. \cite{LM17} in the form of the following theorem: For every state different from the identity we may find a joint measurement so that the state violates the corresponding complementarity inequality, being therefore nonclassical.

\bigskip

To prove this we define the measurement choosing coordinates so that basic observables $\sigma_X$ and $\sigma_Z$ are defined so that $s_X=s_Z= 0$ with $s_n = \AL{s_Y}= |\bm{s}|$ so that the complementarity inequalities are satisfied provided that 
\begin{equation}
\label{fac}
  |\bm{s}| \leq  \frac{\gamma_X \gamma_Z}{\gamma_{XZ}} .
\end{equation}
 Finally, for all $|\bm{s}| \neq 0$ we may always find values of $\gamma_{XZ}$, $\gamma_X$ and $\gamma_Z$  such that (\ref{fac}) is violated. A clear and simple particular example is provided in Sec. IV.

\bigskip

\bigskip

\subsection{Joint probability distribution}

The above inequalities arise by imposing that the distribution $p_\Lambda (x^\prime, z^\prime)$ in Eq. (\ref{plp}) is a legitimate probability distribution. Next we show that this is equivalent to the existence of a legitimate joint probability distribution $p(x,z) \geq 0$ for the incompatible observables obtained by a suitable inversion procedure \cite{AL16a,AL16b,LM17}.

\bigskip

The idea is that the observed marginals $\tilde{p}_X (x)$ and  $\tilde{p}_Z (z)$ contain complete statistical information about $\sigma_X$ and $\sigma_Z$ in the state $\rho$. This is that the actual statistics of $\sigma_X$ in the state $\rho$
\begin{equation}
    p_X (x) =  \frac{1}{2} \left [ 1+ x \mathrm{tr} \left ( \rho  \sigma_X \right ) \right ] ,
\end{equation}
can be obtained from $  \tilde{p}_X (x)$ via a simple inversion procedure 
\begin{equation}
\label{proc}
    p_X (x) = \sum_{x^\prime = \pm 1} \mu_X (x, x^\prime) \tilde{p}_X (x^\prime) ,
\end{equation}
and similarly for $Z$ and $\sigma_Z$, where 
\begin{equation}
\mu_W (w,w^\prime) = \frac{1}{2} \left ( 1 + \frac{w w^\prime}{ \gamma_W}  \right ), \quad w=x,z.
\end{equation}

\bigskip

We can apply the inversion procedure to the complete statistics $\tilde{p} (x,z)$ to get a joint distribution $p(x,z)$ with the correct marginals  for $\sigma_X$ and $\sigma_Z$
\begin{equation}
\label{inv}
    p(x,z)= \sum_{x^\prime, z^\prime = \pm 1} \mu_X (x, x^\prime) \mu_Z (z, z^\prime) \tilde{p} (x^\prime, z^\prime) , 
\end{equation}
leading to 
\begin{equation}
\label{pxz}
    p (x,z) = \frac{1}{4 } \left [ 1+ x \mathrm{tr} \left ( \rho  \sigma_X \right )+  z \mathrm{tr} \left ( \rho  \sigma_Z \right ) + xz \frac{\gamma_{XZ}}{\gamma_X \gamma_Z} \mathrm{tr} \left ( \rho  \sigma_n \right )\right ] .
\end{equation}

\bigskip

What we get is that inverted joint distribution $p(x,z)$ is exactly the same distribution $p_\Lambda (x,z)$ in Eq. (\ref{plp}) as it can be readily seen using Eqs. (\ref{mxmz}) and (\ref{mxz}) .

\bigskip

Therefore, the violation of the complementarity inequalities is equivalent to the lack of a joint probability distribution for the $\sigma_X$ and $\sigma_Z$ observables.  In other words it is equivalent to the nonclassicality of the system state $\rho$, and also equivalent to a non separable joint statistics $\tilde{p}(x,z)$.

\bigskip

\section{Young interferometer example}

Path-interference duality is a classic example of complementarity at work, so it will provide a nice illustration of the above formalism \cite{GBAL18}. Let us represent the state of a Young interferometer by the two orthogonal kets $|\pm \rangle$ that may be regarded as the eigenvectors of the Pauli matrix $\sigma_Z$, so $\sigma_Z$ is the path observable. Interfere occurs by the coherent superposition of $|\pm \rangle$, so we may represent interference by the observable $\sigma_X$.

\bigskip

We will consider that  $\sigma_X$ is directly measured on the system space by projection on its eigenstates $| x\rangle$ that physically would correspond roughly speaking to the detection of the photon on two points of a screen representing maximum and minimum of interference. 

Path information will be transferred from the system space to an auxiliary space. Following a suggestive optical implementation of the interferometer, such auxiliary space may be the polarization of the light at each aperture. The path information can be imprinted in the polarization states by a different phase plate placed on each aperture. When the apertures are illuminated by right-handed circularly polarized light, represented by the vector $|\circlearrowright \rangle$ in the polarization space, the phase plates produce the following aperture-dependent polarization transformation  
\begin{equation}
\label{tfmt}
    |\pm\rangle |\circlearrowright \rangle \rightarrow |\pm\rangle |\pm \theta \rangle ,
\end{equation}
being 
\begin{equation}
    |\pm \theta \rangle = \cos \frac{\theta}{2} |\circlearrowright \rangle \pm \sin \frac{\theta}{2} |\circlearrowleft \rangle .
\end{equation}
Then, the path information is retrieved by measuring any combination of the  observables represented by the Pauli matrices $\Sigma_X$ and $\Sigma_y$ in the auxiliary polarization space spanned by $|\circlearrowright \rangle $ and $ |\circlearrowleft \rangle $ as eigenvectors of $\Sigma_Z$, say 
\begin{equation}
    \Sigma_\varphi = \cos \varphi \Sigma_X - \sin \varphi \Sigma_y ,
\end{equation}
and we denote here Pauli matrices by capital letters to emphasize that they are defined not in the system space but in the auxiliary polarization space. 
This polarization measurement can be very easily achieved in practice with the help of a linear polarizer, where $\varphi$ represents the orientation of its axis. 
\begin{figure}[h]
    \centering
    \includegraphics[width=8 cm]{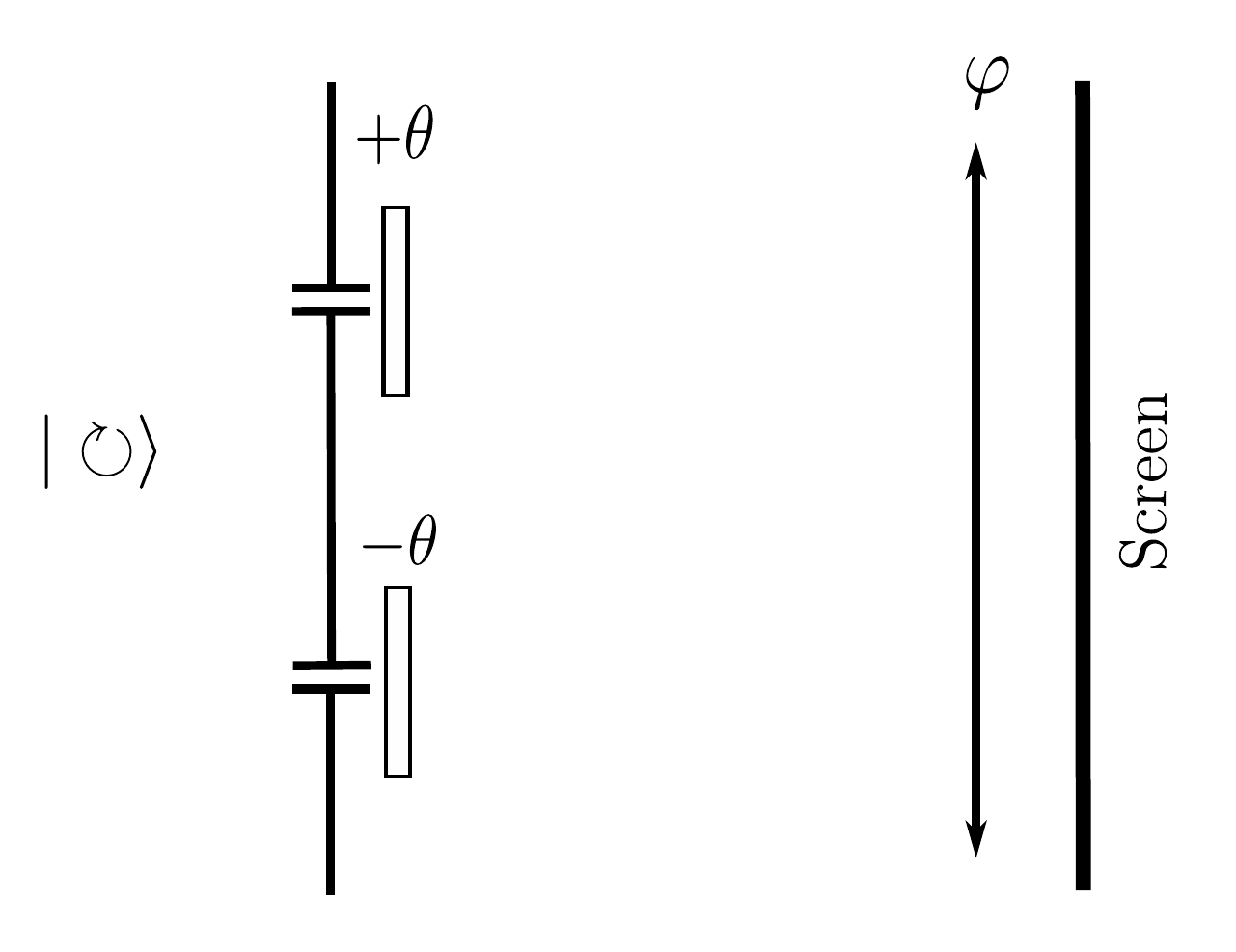}
    \caption{Young interferometer scheme}
    \label{fig:my_label}
\end{figure}

We denote as $|z \rangle_\varphi$ the eigenvectors of  $\Sigma_\varphi$, i. e.,  $\Sigma_\varphi|z \rangle_\varphi = z |z \rangle_\varphi$ with $z=\pm 1$.  The photon passing through the polarizer is represented by the vector $|1 \rangle_\varphi$ while the photon being stopped by the polarizer is represented by the vector $|- 1 \rangle_\varphi$.
 
\bigskip

Thus it can be seen that the statistics of such quantum measurement when the system state is $\rho$ in Eq. (\ref{rho}) leads to the following joint statistics by projection on the system-polarization states $|x\rangle |z \rangle_\varphi$ after the transformation (\ref{tfmt}) 
\begin{equation}
    \tilde{p} (x,z) = \frac{1}{4} \left ( 1+ x \gamma_X s_X + z \gamma_Z s_Z   + xz \gamma_{XZ} s_y \right ) ,
\end{equation}
with 
\begin{equation}
    \gamma_X = \cos \theta ,\quad \gamma_Z = \cos \varphi \sin \theta, \quad \gamma_{XZ}= \sin \varphi \sin \theta ,
\end{equation}
being in this case $\sigma_n = \sigma_y$. Note that the $\gamma$ factors are actually points on the surface of a unit sphere, 
\begin{equation}
    \gamma_X^2+\gamma_Z^2+\gamma_{XZ}^2 = 1 .
\end{equation}

\bigskip

The path observation is more perfect, i, e. $\gamma_Z \rightarrow 1$, as $\varphi \rightarrow 0$ and $\theta \rightarrow \pi/2$ in which case tends to be no observation of interference $\gamma_X \rightarrow 0$, while the interference is more accurate $\gamma_X \rightarrow 1$ as $\theta \rightarrow 0$, in which case tends to be no path observation $\gamma_Z \rightarrow 0$. Moreover, the factor in Eq. (\ref{fac}) becomes
\begin{equation}
 \frac{\gamma_X \gamma_Z}{\gamma_{XZ}} = \frac{\cos \theta}{\tan \varphi} ,
\end{equation}
so that nonclassical results are more clearly revealed as $\theta \rightarrow \pi/2$, $\varphi \rightarrow \pi/2$ so that $\gamma_X \gamma_Z \rightarrow 0$ and $\gamma_{XZ} \rightarrow 1$.

\bigskip

\section{Detection models}

\bigskip

In this section we want to direct a spotlight on the detector, so we will describe it on its most general form, letting the phisical assumptions to the state description. To. this end we express our classical model as
 \begin{equation}
 \label{mvo}
    \tilde{p} (x,z) = \sum_{\lambda^\prime} \tilde{q}(x,z| \lambda^\prime) p_\Lambda (\lambda^\prime ),
\end{equation}
where $\tilde{q}(x,z|\lambda^\prime)$ is the conditional probability describing the whole measurement we do not even assume to factorize. As in the previous model, $p_\Lambda(\lambda)$ describes the probability distribution for the variables $\lambda$, that also will be a pair  of  dichotomic  variables, $\lambda^\prime=  (x^\prime,z^\prime)$ with $x^\prime,z^\prime=\pm1$. Our purpose is to formulate the minimum hypotheses possible about $\tilde{q}(x,z|x^\prime,z^\prime)$.

\bigskip

Sooner or later $\tilde{q}(x,z|\lambda^\prime)$ will be influenced by how $ p_\Lambda (x^\prime,z^\prime )$ is defined, and this is not a trivial question as far as we are dealing with complementary variables. Let us consider the most classical-like distribution compatible with quantum mechanics in the form of a discrete $Q$-like distribution
 \begin{equation}
     p_\Lambda (x^\prime,z^\prime ) = \frac{1}{2} \left [ 1 + \bm{s} \cdot \bm{n} (x^\prime,z^\prime ) \right ] , 
\end{equation}
with
 \begin{equation}
    \bm{n} ( x^\prime,z^\prime ) = \frac{1}{\sqrt{3}} \left ( x,z,xz \right ) ,
\end{equation}
so $ p_\Lambda (x^\prime,z^\prime )$ is always positive and given by projection \AL{of the system density-matrix on four pure states} with $\bm{s} = \bm{n} (x^\prime,z^\prime )$ in  Eq. (\ref{rho}), \AL{which are actually SU(2) coherent states.}

\bigskip

We start then with a general form for $\tilde{q}(x,z|\lambda^\prime)$  based on a Taylor series as done in the previous sections. Our only assumptions link the mean values and correlations of $X$ and $Z$ in the observed statistics, $\overline{x}$, $\overline{z}$, with the mean values of the noiseless observables $\sigma_X$ and $\sigma_Z$, as in (\ref{mxmz}),
\begin{equation}
\label{cond2}
   \overline{x} = \gamma_X  s_x ,  \quad \overline{z} = \gamma_Z s_z.
\end{equation}
As before, $\gamma_X$ and $\gamma_Z$ express the noisy  character  of  the  joint  observation. For definiteness we will consider $\sigma_n = \sigma_Y$, this is to say all the information about the observables $\sigma_{X,Z}$ is already contained in the $x$ and $z$ variables, respectively, so that 
\begin{equation}
\label{cond3}
   \overline{xz}=\gamma_{XZ} s_Y .
\end{equation}
We also take into account that $\tilde{p}(x,z)$ is normalized, since it is a legitimate probability distribution. 

\bigskip

With all of this we obtain the most general form for $\tilde{q}(x,z|x^\prime, z^\prime)$, to be 
\begin{equation}
\label{qt}    
    \tilde{q}(x,z|x^\prime, z^\prime)=\frac{1}{4} \left [1+  \sqrt{3} \left ( \gamma_X x x^\prime  + \gamma_Z z z^\prime +\gamma_{XZ} x z x^\prime z^\prime \right )\right ].
\end{equation}
We may say that $\tilde{q}(x,z|x^\prime, z^\prime)$ is the finite-dimension analog of the Glauber-Sudarshan $P$-function associated to the POVM $\tilde{\Delta} (x,z)$ describing the measurement, $\tilde{p} (x,z) = \mathrm{tr}[\tilde{\Delta}(x,z) \rho]$ \cite{LA20,LP98,AL04,AL06}.

\bigskip

The primary meaning of $\tilde{q}(x,z|x^\prime, z^\prime)$ is a conditional probability that would require $\tilde{q}(x,z|x^\prime, z^\prime)\geq 0$ for all $x, z, x^\prime, z^\prime$, that leads to the following inequalities:
\begin{equation}
\label{Bq}
   1-\sqrt{3}|\gamma_X-\gamma_Z|\geq \sqrt{3} \gamma_{XZ}\geq \sqrt{3}|\gamma_X+\gamma_Z|-1 .
\end{equation}
Moreover, $\tilde{q}(x,z|x^\prime, z^\prime)$ admits a factorization of the form
\begin{equation}
    \tilde{q}(x,z|x^\prime, z^\prime)=\tilde{q}(x|x^\prime)\tilde{q}(z| z^\prime)
\end{equation}
if and only if
\begin{equation}
    \label{Qfac}
    \gamma_{XZ}=\sqrt{3} \gamma_X\gamma_Z.
\end{equation}

\bigskip

So, we may say that violation of any of these conditions may be regarded as reflecting the quantum nature of the observation process. Furthermore if we combine both we get $\tilde{q}(x,z|x^\prime, z^\prime) \geq 0$ for the factorized case if and only if
\begin{equation}
\label{Pp}
    \gamma_X \leq \frac{1}{\sqrt{3}}, \quad \gamma_X \leq \frac{1}{\sqrt{3}}.
\end{equation}

\bigskip

\begin{figure}[h]
    \centering
    \includegraphics[width=6cm]{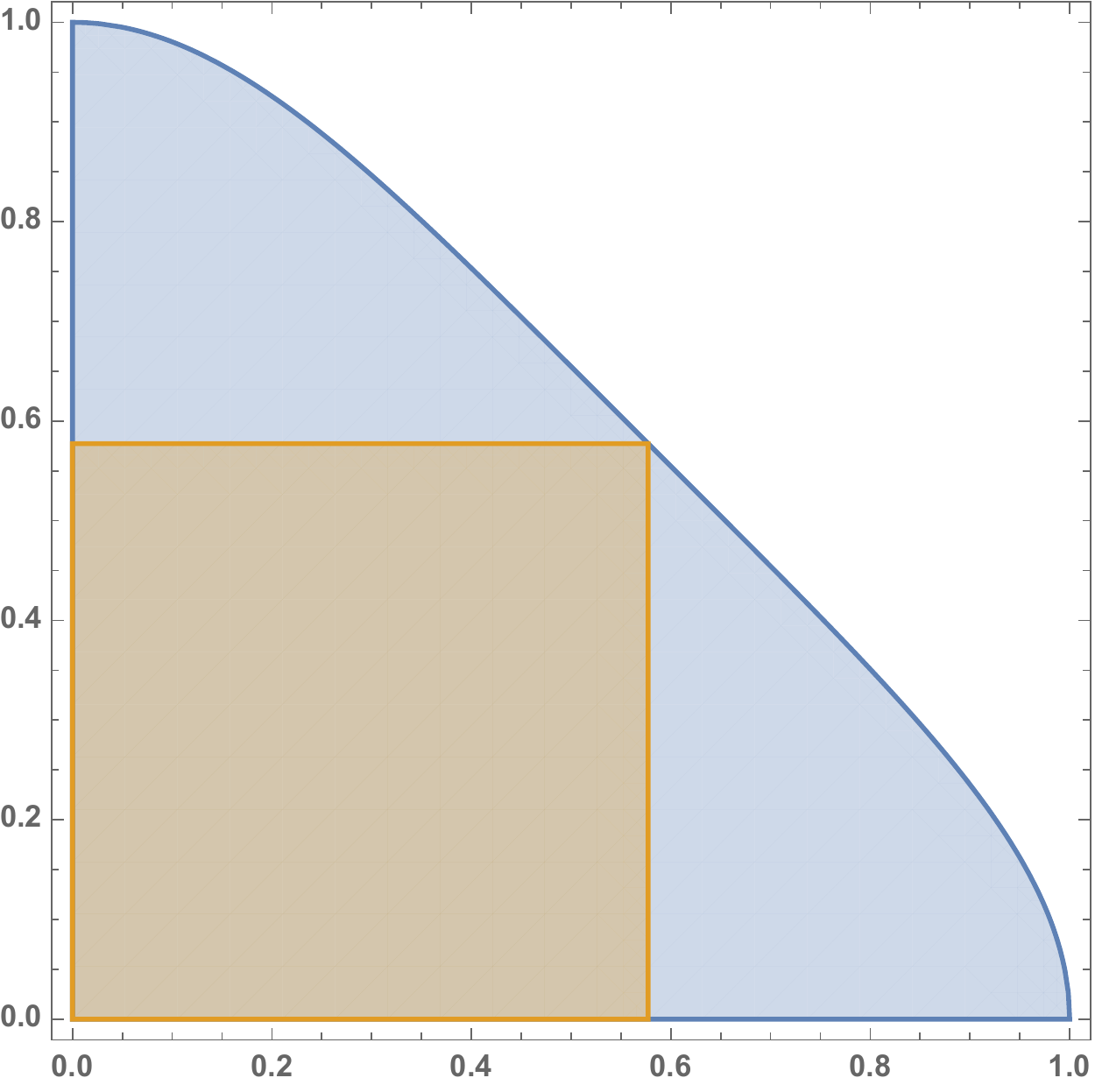}
    \caption{Regions in the $(\gamma_X,\gamma_Z)$ plane satisfying $\tilde{\Delta} (x,z) \geq 0$ in Eq. (\ref{tdpf}) as the region limited by the blue line, as well as the square region satisfying the nonnegativity of the factorized $\tilde{q}(x,z|x^\prime, z^\prime)$ in Eq. (\ref{Pp}).}
   \label{eta1}
\end{figure}

Let us recall that $\tilde{\Delta} (x,z) \geq 0$ requires that
\begin{equation}
    \gamma_X^2+\gamma_Z^2+\gamma_{XZ}^2 \leq 1 ,
\end{equation}
that along with factorization becomes 
\begin{equation}
\label{tdpf}
    \gamma_X^2+\gamma_Z^2+3 \gamma_X^2 \gamma_Z^2 \leq 1 .
\end{equation}
In Fig. \ref{eta1} we show the region in the $(\gamma_X,\gamma_Z)$ plane satisfying $\tilde{\Delta} (x,z) \geq 0$ in Eq. (\ref{tdpf}) as the region limited by the blue line, as well as the square region satisfying the nonnegativity of the factorized $\tilde{q}(x,z|x^\prime, z^\prime)$ in Eq. (\ref{Pp}).
{}

\bigskip

Next, we apply the inversion procedure introduced in Sec.III to the conditional probability $\tilde{q}(x,z|x^\prime, z^\prime)$ 
\begin{equation}
q(x,z|x^{\prime\prime},z^{\prime\prime})=\sum_{x^\prime,z^\prime} \mu_X (x, x^\prime) \mu_Z (z, z^\prime)  \tilde{q}( x^\prime, z^\prime | x^{\prime \prime}, z^{\prime \prime}) ,
\end{equation}
where $\mu_W(w,w^\prime)$ $w=x,z$ are the same functions defined in (\ref{inv}), to get
\begin{equation}
q(x,z|x^{\prime\prime},z^{\prime\prime})=\frac{1}{4} \left[ 1+ \sqrt{3} \left ( xx^{\prime\prime}+zz^{\prime\prime}+\frac{\gamma_{XZ}}{\gamma_X \gamma_Z} xzx^{\prime\prime}z^{\prime\prime}\right) \right ].
\end{equation}
Then, nonnegativity $q(x,z|x^{\prime\prime},z^{\prime\prime}) \geq 0$ leads to the following inequalities:
\begin{equation}
\label{Bqg}
   1 \geq \sqrt{3} \frac{\gamma_{XZ}}{\gamma_X \gamma_Z}\geq 2\sqrt{3}-1 ,
\end{equation}
that can never be satisfied as far as $2\sqrt{3}-1 >1$. On the other hand, the condition of factorization $\tilde{q}(x,z|x^{\prime\prime},z^{\prime\prime})=\tilde{q}(x|x^{\prime \prime})\tilde{q}(z| z^{\prime \prime})$ is again the same in Eq. (\ref{Qfac}). 

\bigskip

\section{Conclusions}
We have provided a complete analysis of complementarity including a classical model and practical complementarity inequalities whose violation is fully equivalent to a lack of a joint distribution for the corresponding incompatible observables. All this is clearly illustrated by path-interference duality in a Young interferometer.

\bigskip

\AL{It is clear that there are many parallels between our complementarity formalism and the standard approach used to derive Bell inequalities in a bipartite setting.} \LA{The establishment of the classical separable model in Eq.(\ref{Bt}) formally adopts the standard assumptions for the derivation of Bell-like hidden variables models \cite{MJWH15,BD21}}.\AL{The main difference is that in our case there is no locality issues related to Eq. (\ref{Bt}), since there are not two parties but just one, and entanglement plays no role.} \LA{In the case of Bell's theorem, the separability is required between the subsystems' conditional probabilities, while in this work we apply this idea of statistical independence to the conditional probability of each individual observable. Naturally, the approach of the fuzzy joint measurement of incompatible observables can be extended to an entangled bipartite system so that the complete Bell inequalities can be recovered \cite{MAL20}. Then, we can also extend this study of complementarity to the complete Bell scenario by considering the factorization of the conditional probabilities not only between subsystems but also between observables within the same subsystem. Thus, in this complete Bell scenario, our approach to complementarity may be a fruitful tool to disengage the roles of complementarity and entanglement \cite{BD21}. }

\bigskip

\AL{Because of this, given its independence of non-locality and entanglement issues, our formalism with the separable classical-like model in Eq. (\ref{Bt}) might be rather linked to the investigation of methods to detect quantum contextuality \cite{AC08,KZGKGCBR09}, specially since it has been questioned the proper relation between contextuality and  incompatibility \cite{AC21,AK21b,SSWSKS21}. In this regard it is worth noting that we have shown in Sec. 3.4 that every state different from the identity can violate the classical-like inequalities for a properly chosen experimental setting, so at difference with Bell-type inequalities these hold for all quantum states except the trivial identity. Thus, like contextuality, this points to a very basic difference between the classical and quantum theories, and this is that classical observed statistics are always separable in the sense of Eq. (\ref{Bt}), while for every quantum system we can find measurements with non separable statistics. Finally, we find it valuable that this approach addresses the studied quantum features in terms of practical feasible observations instead of abstract definitions in Hilbert spaces.   }

\bigskip

\section*{ACKNOWLEDGMENTS}
L. A. and A. L. acknowledge financial support from Spanish Ministerio de Econom\'ia y Competitividad Project No. FIS2016-75199-P. 
L. A. acknowledges financial support from European Social Fund and the Spanish Ministerio de Ciencia Innovaci\'{o}n y Universidades, Contract Grant No. BES--2017--081942.

\end{document}